\documentclass[12pt]{iopart}

\usepackage{iopams}  
\usepackage{bbm}
\usepackage{graphicx}  

\newcommand{\eps}{\varepsilon}

\begin{document}

\title{Optimal trapping of monochromatic light in designed photonic multilayer structures}

\author{Fabian Spallek, Andreas Buchleitner and Thomas Wellens}
\address{Albert-Ludwigs-Universit{\"a}t Freiburg, Physikalisches Institut, Hermann-Herder-Str. 3, D-79104 Freiburg i. Br.}

\date{\today}

\begin{abstract}
We devise an optimised bi-component multi-layered dielectric stack design to enhance the local irradiance for efficient photovoltaic upconversion materials. 
The field intensity profile throughout the photonic structure is numerically optimized by appropriate tuning of the individual layers' thicknesses.
The optimality of the thus inferred structure is demonstrated by comparison with an analytically derived upper bound.
The optimized local irradiance is found to increase exponentially with the number of layers, its rate determined by the permittivity ratio of the two material components.
Manufacturing errors which induce deviations from the optimised design are accounted for statistically, and set a finite limit to the achievable enhancement.
Notwithstanding, realistic assumptions on manufacturing errors still suggest achievable irradiances which are significantly larger than those obtained with the recently proposed Bragg stack structures.

\end{abstract}

\maketitle

\section{Introduction}
\label{sec:Introduction}
Photon upconversion, i.e., the conversion of two photons with smaller energy into one photon with larger energy, 
offers promising possibilities to improve the efficiency of solar cells \cite{brown_third_2009, sark_upconversion_2013, trupke_improving_2002}. The majority of currently used photovoltaic technologies is based on materials with semiconductor properties, where a certain minimum energy per photon (defined by the bandgap of the semiconductor) is required to create an electron-hole pair. Photons with smaller energies are not absorbed by the solar cell, and are therefore lost for the purpose of light-energy conversion. In the case of silicon, the resulting loss amounts to approximately $20\%$ of the power of the incident solar radiation \cite{richards_enhancing_2006}.

In principle, these losses can be reduced by shifting the energy of the transmitted photons above the bandgap, thereby rendering them usable for the solar cell. Experimentally, a thus achieved relative increase of the current generated 
by a silicon solar cell of about $0.55\%$ has already been demonstrated \cite{fischer_record_2015}.
This increase was realised by placing an upconverter material made of monocrystalline ${\rm BaY}_2{\rm F}_8:30\%~{\rm Er}^{3+}$ on the non-irradiated side of the solar cell. 
In order to further increase the upconversion efficiency, it was proposed to embed the upconversion material within suitably chosen photonic structures
\cite{herter_increased_2013, hofmann_upconversion_2016}. 
In principle, such structures may influence the upconversion luminescence in different ways: first, they can be used in order to enhance the local irradiance of the upconversion material as compared to the non-concentrated 
incident light. 
This enhancement also increases the upconversion luminescence due to the nonlinearity of the underlying process \cite{pollnau_power_2000}. 
Second, photonic structures change the local density of states and, consequently \cite{agarwal_quantum_1975}, the rates of spontaneous emission processes occurring within the upconverter ions. Recently, a Bragg stack consisting of layers with periodically alternating refractive indices has been identified as a particularly promising candidate of a photonic structure, which, according to the  results of numerical simulations, leads to a considerable enhancement of the upconversion luminescence for experimentally realistic parameters \cite{hofmann_upconversion_2016}.

According to the design proposed in \cite{hofmann_upconversion_2016}, the thicknesses of all layers forming the Bragg stack are determined by a single parameter (the so-called design wavelength $d$, see section~\ref{subsec:ResultsOptimization} below), which is then optimized in order to maximize the upconversion luminescence. 
In the present paper, we study to which extent such structures can be further improved by optimizing the thickness of each constituent layer individually, such as to maximize the local irradiance at a given position. 
We design a photonic structure consisting of dielectric layers with given permittivities such that, for an incident monochromatic plane wave, the intensity of the electric field inside the photonic structure is maximized. In other words, we aim at trapping the incident photons inside the photonic structure for a time as long as possible, by the creation of strongly localized, narrow resonance eigenmodes which result from the boundary conditions as set by the permittivity landscape. 
In contrast to traditional methods of light trapping \cite{redfield_multiplepass_1974, yablonovitch_statistical_1982} based on principles of geometric optics, our optimization strategy relies on appropriately tuned interferences between multiply reflected wave amplitudes, and is therefore restricted to waves with a given wavelength and angle of incidence.
From a more general perspective, our problem to define the optimal photonic structure is closely related to problems of photon management \cite{vynck_photon_2012} and of network design for optimised excitation transport in light harvesting units made by nature \cite{scholak_efficient_2011, scholak_ch.1_2010, walschaers_quantum_2016, walschaers_scattering_2017}

The paper is organized as follows: in section~\ref{sec:Methods}, we introduce the Helmholtz equation for the electric field at a given position inside a one-dimensional multi-layered photonic structure, describe the transfer matrix method  as an efficient tool for its solution and outline the numerical algorithm to determine optimized structures in which the intensity of the electric field 
is maximized. In section~\ref{sec:Results}, we derive analytical upper bounds for the achievable intensity enhancement and compare them to the characteristic properties of our numerically optimized structures. In section~\ref{sec:Conclusion}, we summarize and discuss our results, and give perspectives for future studies.

\section{Model and Methods}
\label{sec:Methods}

We consider a stationary scattering scenario for a light field of given polarization, fixed frequency $\omega$, and wave vector normal to the photonic structure's surface. The time dependence of the electric field is separated as
\begin{equation}
\vec{E}(\vec{r},t) = \vec{E}_{\omega}(\vec{r}) \, e^{-i \omega t} \;,
\end{equation}
and Maxwell's equations (without sources) can be condensed into the Helmholtz equation for the space-dependent electric field component $\vec{E}_{\omega}(\vec{r})$:
\begin{equation}\nabla^2 \vec{E}_{\omega} + \frac{\omega^2}{c^2} \, \eps(\vec{r}) \vec{E}_{\omega} = 0 \; . 
\label{eq:helmholtz}   
\end{equation}
Given a one-dimensional multilayer structure with layers in the $x-y$ plane, and homogeneous dielectric materials defined by a piecewise constant permittivity $\eps(\vec{r})=\eps(z)=\eps_n$ in the $n$-th layer, without absorption, i.e., $\eps(z)\in{\mathbbm R}$,
we are left with the one-dimensional Helmholtz equation: 
\begin{equation}\label{eq:HarmOsci}
\frac{\partial^2}{\partial z^2} E_{\omega} + \frac{\omega^2}{c^2} \, \eps_n E_{\omega} = 0
\end{equation}
in layer $n \in \{1,\ldots, N\}$.
The general solution of (\ref{eq:HarmOsci}) reads
$E_{\omega}(z)=A_n e^{ik_n z}+B_n e^{-ik_n z}$, with wave number $k_n=\omega \sqrt{\eps_n}/c$ and
amplitudes $A_n$ (and $B_n$) for right- (and left-) propagating waves.

\begin{figure}
	\centering
	\includegraphics[width=0.50\textwidth]{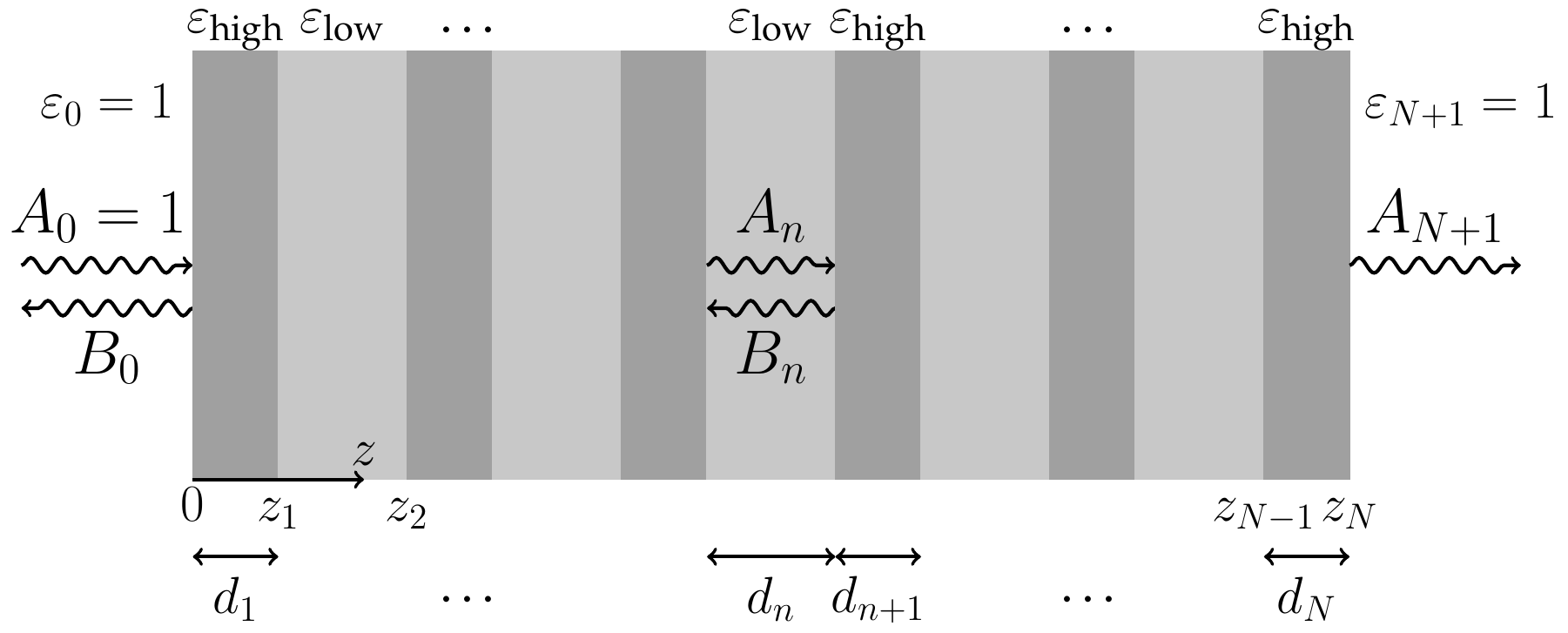}
	\centering
	\caption{Photonic structure consisting of alternating layers with permittivities $\eps_{\rm high}$ and $\eps_{\rm low}$, respectively. The amplitudes of the right- and left-propagating waves inside layer $n$ (with thickness $d_n=z_n-z_{n-1}$) are denoted by $A_n$ and $B_n$, respectively. The incident wave with normalized amplitude $A_0=1$ arrives from the left-hand side.}
	\label{fig:SketchGeo}
\end{figure}

The multilayer structure consists of two alternating materials, one with permittivity $\eps_{{\rm low}}$ containing the upconverter, and a second one with larger permittivity $\eps_{{\rm high}}>\eps_{{\rm low}}$ providing the photonic structure.
We consider an odd number $N$ of layers, with permittivities $\eps_n=\eps_{{\rm high}}$ for odd, and $\eps_n=\eps_{{\rm low}}$ for even $n$, respectively. The $n$-th layer is confined between  $z_{n-1}$ and $z_n$, with $z_n > z_{n-1}$ and $z_0=0$. The stack is embedded in air, such that $\eps_0=1$ for $z<0$ and $\eps_{N+1}=1$ for $z > z_N$.
The stack is thus defined by the layer thicknesses  $d_n=z_n-z_{n-1}$ and the associated values of $\eps_{{\rm high}}$ and $\eps_{{\rm low}}$, see figure  \ref{fig:SketchGeo}.

\subsection{Transfer matrix method}
\label{subsec:MethodsTransfermatrix}
The general solution of (\ref{eq:helmholtz})
for a one-dimensional structure
as depicted in figure \ref{fig:SketchGeo} reads:
\begin{eqnarray}
E_{\omega}(z) & = & \sum_{n=1}^N \left( A_{n} e^{i k_{n} z} + B_{n} e^{- i k_{n} z } \right) \, \Theta(z-z_{n-1}) \, \Theta(z_n-z) \nonumber\\
& + & \left( A_{0} e^{i k_{0} z} + B_{0} e^{- i k_{0} z } \right) \, \Theta(-z) \nonumber\\
& + & \left( A_{N+1} e^{i k_{N+1} z} + B_{N+1} e^{- i k_{N+1} z } \right) \, \Theta(z-z_{N})
\label{eq:FullE} \;,
\end{eqnarray}
with amplitudes $A_n$ and $B_n$ related through the boundary conditions at the layers' interfaces.
Maxwell equations in non-magnetic media without external sources enforce the continuity of $E_{\omega}$ as well as of $\partial_z E_{\omega}$ 
at the position $z_n$ of every interface, i.e.: 
\begin{eqnarray}
A_{n} e^{i k_{n} z_{n}} + B_{n} e^{- i k_{n} z_{n}} &= & A_{n+1} e^{i k_{n+1} z_n} + B_{n+1} e^{- i k_{n+1} z_n} \;,\\
k_n A_{n} e^{i k_{n} z_{n}} - k_n B_{n} e^{- i k_{n} z_{n}} &= & k_{n+1} A_{n+1} e^{i k_{n+1} z_n} - k_{n+1} B_{n+1} e^{- i k_{n+1} z_n} \;.
\end{eqnarray} 
This relates the amplitudes $v_n=(A_n, B_n)^T$ and $v_{n+1}=(A_{n+1}, B_{n+1})^T$  within two adjacent layers $n$ and $n+1$ according to
\begin{equation}
v_{n}= \textbf{M}_{n} \cdot v_{n+1} \;,
\label{eq:transfer}
\end{equation}
with the transfer matrix \cite{abeles_notitle_1950}
 \begin{equation}
 \textbf{M}_{n} = \frac{1}{2 \, k_{n}}
 \left(\begin{array}{cc}
 \left(k_{n} + k_{n+1}\right)  e^{i (k_{n+1}-k_{n}) z_{n} } &  \left(k_{n} - k_{n+1}\right)  e^{- i (k_{n+1}+k_{n}) z_{n} } \\
 \left( k_{n} - k_{n+1} \right)  e^{i (k_{n+1}+k_{n}) z_{n} } &  \left(k_{n} + k_{n+1}\right)  e^{- i (k_{n+1}-k_{n}) z_{n} } 
 \end{array}
 \right) \;.
 \label{eq:tmatrix}
 \end{equation}
Note that $\textbf{M}_{n}$ is fully determined by the geometry encoded in $\eps(z)$, as it only depends on the positions $z_n$ of the surfaces between the layers, and on the wave vectors $k_n = \omega\sqrt{\eps_n}/c$.
The amplitude $A_0$ of the incoming wave is normalized such that $A_0=1$.
Because of possible reflection at $z_0 = 0$, it follows that $B_0 \neq 0$ and, since there is no incoming wave travelling in negative $z$-direction from $z>z_N$, we have $B_{N+1}=0$.

The reflection coefficient of the full structure is given by 
$R=|B_{0}|^2$
while the transmission reads  
$T=|A_{N+1}|^2$. From energy flux conservation, it follows that $R+T=1$ and
\begin{equation}\label{eq:fluxConserv}
\sqrt{\eps_{n}} \left( |A_{n}|^2 - |B_{n}|^2 \right) = T 
\end{equation}
for each layer $n$.

The solution of (\ref{eq:FullE}), given by the amplitudes $v_n=(A_n, B_n)^T$, $n=1, \ldots, N$, can now be determined by propagation of the initial condition $(\tilde{A}_{N+1}, \tilde{B}_{N+1}) = (1,0)$ across the multilayer structure, by iterative application of the transfer matrix $\textbf{M}_{n}$: $(\tilde{A}_n, \tilde{B}_n)^T= \textbf{M}_{n} \cdot (\tilde{A}_{n+1}, \tilde{B}_{n+1})^T$. 
Once the input face of the structure is reached, $(\tilde{A}_0, \tilde{B}_0)^T= \textbf{M}_0\cdot \textbf{M}_1\cdot\dots \cdot \textbf{M}_N\cdot (1, 0)^T$, in a final step all amplitudes $\tilde{A}_n$, $\tilde{B}_n$ are renormalized, i.e. $A_n=\frac{\tilde{A}_n}{\tilde{A}_0}$ and $B_n=\frac{\tilde{B}_n}{\tilde{A}_0}$, such that $A_0=1$.

\subsection{Numerical optimization}
\label{subsec:MethodsOptimization}

To improve the efficiency of the upconversion process, we seek to maximize the field intensity within the upconversion layers of the photonic structure. The corresponding target function is given by the intensity enhancement factor
\begin{equation}
\gamma =   \frac{1}{\sum_{m=1}^{\frac{N-1}{2}} (z_{2m}-z_{2m-1})}
\sum_{m=1}^{\frac{N-1}{2}} \int_{z_{2m-1}}^{z_{2m}}  \mathrm{d}z \, I(z) 
\;\;,
\label{eq:target}
\end{equation}
which quantifies the average intensity in the upconversion volume of the multilayer device, and owes its name to the fact that $I$ is normalized to the intensity of the incoming wave.
At least at weak irradiation, the upconversion efficiency is expected to scale quadratically with the intensity \cite{pollnau_power_2000}. One may therefore alternatively average over $I^2(z)$ in (\ref{eq:target}) rather than over $I(z)$. However, we verified that our subsequent results are essentially insensitive to such a replacement.

Note that, due to the periodic intensity modulation within each layer $n$ given by
\begin{eqnarray}
I(z) &= & |A_n|^2+|B_n|^2 + 2 \; \Re \left( A_n \, B_n^* \; e^{2 i k_n z} \right) \nonumber\\
&=  & |A_n|^2+|B_n|^2 + 2 |A_n| |B_n| \cos \left( 2 k_n z + \varphi_n \right)
\label{eq:IntensityLayerAB} \;,
\end{eqnarray}
which is a direct consequence of the superposition of right- and left-running amplitudes $A_n$ and $B_n$, respectively, it suffices to optimize $d_n$ in the range $d_{\rm min} < d_n < \frac{\lambda_n}{2} - d_{\rm min}$, with $\lambda_n=2\pi/k_n$ and $d_{\rm min}= 0.025 \lambda_n$ to exclude vanishing layer thicknesses as output of our optimization procedure.

The optimal thickness profile $d_n, n=1,\ldots,N$, is then numerically inferred through the following iterative procedure:
\begin{enumerate}
	\item \label{list:insertLayers} Insert two additional layers with thickness $d_{\frac{M-3}{2}+1} = d_{\frac{M-3}{2}+2} = \frac{\lambda_n}{2} $ in the middle of the stack of a given optimized structure with $M-2$ layers, with these previously given $M-2$ layers' thicknesses unchanged.
	\item \label{list:downhillsimplex} Optimize this $M$-layer structure with the downhill simplex method of Nelder and Mead \cite{press_numerical_2007, nelder_simplex_1965}.
	\item From the analytical benchmark (\ref{eq:gammaGrow}), we expect that $\gamma$ increases exponentially as a function of $M$, i.e. $\gamma(M)=[\gamma(M-2)]^2/\gamma(M-4)$.
	Verify whether the result of (\ref{list:downhillsimplex}) leads to an enhancement $\gamma$ which agrees 
	(up to a relative error not larger than $2\%$)
	with this expectation.
	\begin{itemize}
	\item If agreement is given, continue with (\ref{list:insertLayers}) until $M=N$. 
	\item If not, re-initialize the downhill simplex algorithm with statistically perturbated initiated values of  $d_{\frac{M-3}{2}+1}$ and $d_{\frac{M-3}{2}+2}$.
	\end{itemize}
\end{enumerate}

\section{Results}
\label{sec:Results}

Before we discuss specific properties of the numerically optimized structures, 
let us first derive an analytical upper bound for the achievable maximal local intensity $I(z)$ at a given position $z$ within the photonic structure, given the fundamental set of equations (\ref{eq:FullE})-(\ref{eq:IntensityLayerAB}). As we will see, the upper bound for $I(z)$ increases exponentially as a function of the number of layers which separate the position $z$ from the left or the right boundary of the photonic structure, respectively. 
From this, we estimate an upper bound for the integrated local intensity, which defines the intensity enhancement factor $\gamma$ as given in Eq.~(\ref{eq:target}). Also $\gamma$ increases exponentially as a function of the total number $N$ of layers.
The above bound enters as a benchmark into the numerical optimization procedure as described in \ref{subsec:MethodsOptimization} further up, and in addition provides useful physical insight.

\subsection{Analytical upper bound}
\label{subsec:ResultsUpperBoundary}

To derive an upper bound for the intensity that can, in principle, be reached with a given number $N$ of layers, we first observe that, according to (\ref{eq:IntensityLayerAB}), 
the intensity $I(z)$ inside layer $n$ fulfills:
\begin{equation}\label{eq:A+B=2I}
I(z)\leq \left( |A_{n}|+|B_{n}| \right)^2=I_n+\sqrt{I_n^2-\frac{T^2}{\eps_n}} \;,
\label{eq:Izbound}
\end{equation}
where we used flux conservation, see (\ref{eq:fluxConserv}), and introduced the background intensity
\begin{equation}
I_n=|A_n|^2+|B_n|^2
\label{eq:Indefinition}
\end{equation}
inside layer $n$, defined as the constant (i.e. non-oscillating) term in the expression (\ref{eq:IntensityLayerAB}) for $I(z)$.
Since $v_n=(A_n,B_n)^T$, the background intensity $I_n$ can be interpreted as the squared norm of the vector 
$v_n\in{\mathbbm C}^2$ induced by the standard scalar product in ${\mathbbm C}^2$, i.e., $I_n=v_n^\dagger\cdot v_n$.
According to (\ref{eq:transfer}), we have:
\begin{equation}
v_{n}^{\dagger} \cdot v_{n} = v_{n+1}^{\dagger} \cdot \textbf{M}_{n}^{\dagger} \cdot \textbf{M}_{n}  \cdot v_{n+1} \;.
\end{equation}
Therefore, the ratio $I_n/I_{n+1}=\left(v_n^\dagger \cdot v_n\right)/
\left(v_{n+1}^\dagger\cdot v_{n+1}\right)$ of the background intensities in adjacent layers is bounded between the two eigenvalues of the 
Hermitian matrix:
\begin{equation}
\textbf{M}_{n}^{\dagger} \cdot \textbf{M}_{n} = \frac{1}{2 \, k_{n}^{2}}
\left(
\begin{array}{cc}
k_{n}^2 + k_{n+1}^2  &  \left(k_{n}^2 - k_{n+1}^2\right)  e^{- 2i  k_{n+1} z_{n} } \\
\left( k_{n}^2 - k_{n+1}^2 \right)  e^{2i k_{n+1} z_{n} } &  k_{n}^2 + k_{n+1}^2 
\end{array}
\right)
\end{equation}
The eigenvalues and eigenvectors of this matrix are:
\begin{eqnarray}
\lambda_n^{(1)}  = 1, & \ \ \ \ & w_n^{(1)} = \frac{1}{\sqrt{2}} \left( 1, e^{2 i k_{n+1} z_{n}} \right) \nonumber \\
\lambda_n^{(2)}  = \left( \frac{k_{n+1}}{k_{n}} \right)^2, & & w_n^{(2)} = \frac{1}{\sqrt{2}} \left( 1, - e^{2 i k_{n+1} z_{n}} \right)
\label{eq:Eigensys}
\end{eqnarray}
It follows that $I_{n+1} k_{n+1}^2/k_n^2 \leq I_n\leq I_{n+1}$ if $k_n\geq k_{n+1}$ and $I_{n+1} \leq I_n\leq I_{n+1} k_{n+1}^2/k_n^2$ if $k_n\leq k_{n+1}$. In particular, the background intensity in a layer with {\em smaller} permittivity is always {\em larger} than in an adjacent layer with higher permittivity.

An upper bound for the background intensities in each layer can now be obtained as follows: let us consider a scattering process with reflection coefficient $R=1-T$. Due to the boundary conditions mentioned above (no incoming wave from the right-hand side), we have $|A_{N+1}|=\sqrt{T}$ and $B_{N+1}=0$, hence $I_{N+1}=T=1-R$. The amplitudes $|A_N|$ and $|B_N|$ inside the rightmost layer then follow from
(\ref{eq:transfer}) and (\ref{eq:tmatrix}), 
with $k_N=\sqrt{\epsilon_{\rm high}}$, and $k_{N+1}=1$ and
resulting background intensity:
\begin{equation}
I_N=\frac{1-R}{2}\left(1+\frac{1}{\epsilon_{\rm high}}\right)\label{eq:IN}
\end{equation}
The background intensity $I_{N-1}$ in the adjacent layer with permittivity $\epsilon_{\rm low}$ fulfills
$I_{N-1}\leq \alpha I_N$, where
\begin{equation}
\alpha=\frac{\epsilon_{{\rm high}}}{\epsilon_{{\rm low}}}>1
\label{eq:alpha}
\end{equation} 
defines the largest eigenvalue $\lambda_{N-1}^{(2)}$ of $\textbf{M}_{N-1}^\dagger\cdot \textbf{M}_{N-1}$ (with $k_{N}=\sqrt{\epsilon_{\rm high}}$ and $k_{N-1}=\sqrt{\epsilon_{\rm low}}$). In contrast, the eigenvalues of the matrix 
$\textbf{M}_{N-2}^\dagger\cdot \textbf{M}_{N-2}$ corresponding to the following layer (with $k_{N-1}=\sqrt{\epsilon_{\rm low}}$ and $k_{N-2}=\sqrt{\epsilon_{\rm high}}$) are given by $\lambda_{N-2}^{(1)}=1$ and $\lambda_{N-2}^{(2)}=1/\alpha<1$, such that 
$I_{N-2}\leq I_{N-1}$. For each double layer, the background intensity thereby  increases at most by a factor $\alpha$. In total, we obtain $I_n\leq I_n^{({\rm max},r)}$, with upper bound
\begin{equation}
I_n^{({\rm max},r)}=\alpha^{\lfloor\frac{N+1-n}{2}\rfloor} I_N \label{eq:boundright}
\end{equation}
where
$\lfloor\dots\rfloor$ denotes the floor function, i.e. $\lfloor n/2\rfloor = n/2$ if $n$ is even, and $\lfloor n/2\rfloor = (n-1)/2$ if $n$ is odd.

A similar bound can also be derived starting from the left hand side: $A_0=1$ and $B_0=e^{i\phi}\sqrt{R}$. Again, $A_1$ and $B_1$ follow from (\ref{eq:transfer}) and (\ref{eq:tmatrix}), and the maximization over the phase $\phi$ yields:
\begin{equation}
I_1^{({\rm max})} = \frac{\left(1-\sqrt{R}\right)^2+\epsilon_{\rm high} \left(1+\sqrt{R}\right)^2}{2\epsilon_{\rm high}}\label{eq:I1}
\end{equation}
Taking once again into account the maximum amplification factor $\alpha$ for each double layer, we obtain another upper bound for $I_n$,
i.e. $I_n\leq I_n^{({\rm max},l)}$ with
\begin{equation}
I_n^{({\rm max},l)}=\alpha^{\lfloor\frac{n}{2}\rfloor} I_1^{({\rm max})} \label{eq:boundleft} \;.
\end{equation}

Thus, we obtain two upper bounds, one exponentially increasing from right to left, and one exponentially increasing from left to right. The maximal reachable background intensity within the $n$-th layer consequently is
\begin{equation}
I^{({\rm max})}_n = \min \left( I_n^{({\rm max},l)} , I_n^{({\rm max},r)} \right) \label{eq:IntensityMax_n} \;\;.
\end{equation}
The largest intensity which satisfies both bounds occurs at the point where the  two exponential functions intersect each other. Neglecting the floor functions in (\ref{eq:boundright}) and (\ref{eq:boundleft}), this point is found at the layer
$n_{\rm max}=(1+N)/2-\ln(I_1^{({\rm max})}/I_N)/\ln(\alpha)$, with corresponding intensity 
\begin{equation}
I^{({\rm max})}=\sqrt{I_1^{({\rm max})} I_N}~\alpha^{(1+N)/4}\label{eq:IntensityMax} \;\;.
\end{equation}
This intensity can be further optimized by using (\ref{eq:IN}) and (\ref{eq:I1}) to obtain 
$I_1^{({\rm max})}$ and $I_N$ as a function of $R$, and determining the value of $R$ which maximizes the geometric mean
$\sqrt{I_1^{({\rm max})} I_N}$. For example, using $\epsilon_{\rm high}=4$, we find $R=0.219$ as the optimal value of the reflection coefficient.

\begin{figure}
	\centering
			\includegraphics[width=\textwidth]{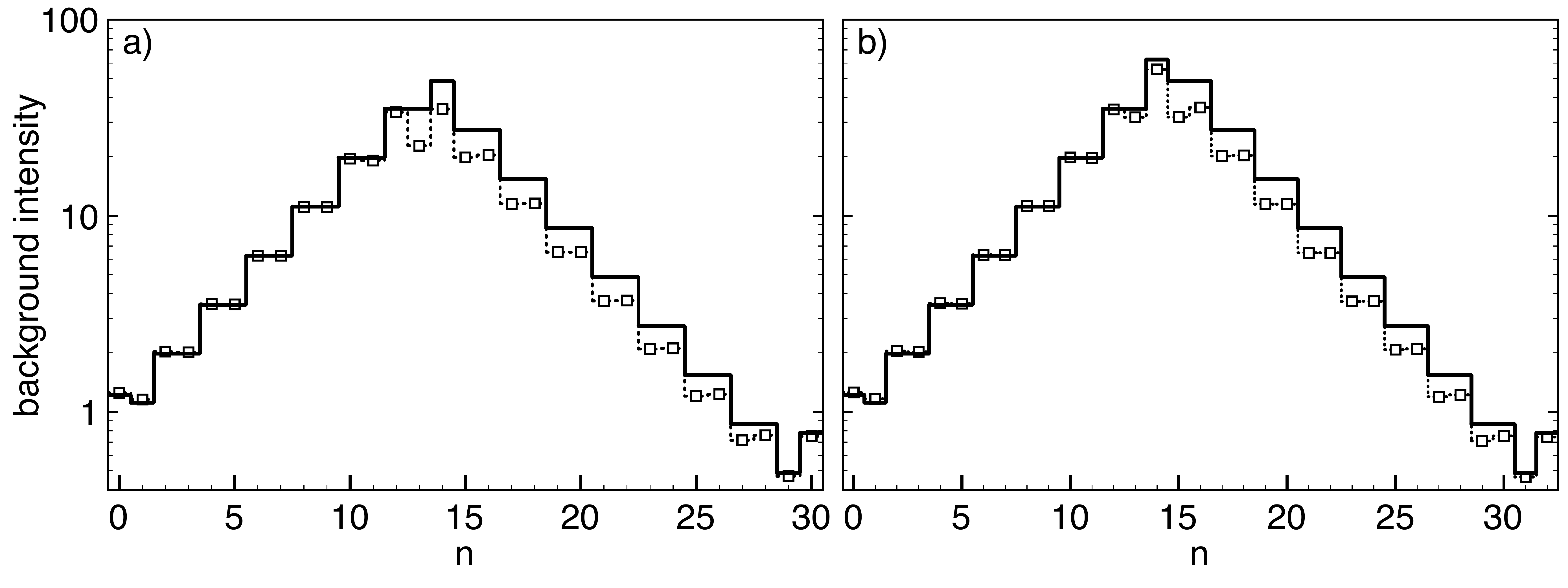}
	\caption{
	Comparison of the background intensities $I_n=|A_n|^2+|B_n|^2$ within the $n$-th layer in the numerically optimized stack (dotted lines with symbols) to the analytically given upper bound (\ref{eq:IntensityMax_n}) (solid lines) for (a) $N=29$ layers, and (b) $N=31$ layers, with $\epsilon_{\rm high}=4$ and $\epsilon_{\rm low}=2.25$. 
	Starting from the left- and rightmost layer ($n=0$ and $n=N+1$), respectively, where $I_0=1+R$ and $I_{N+1}=1-R$ (with reflection coefficient $R$), respectively, the intensity increases exponentially and assumes its maximum value close to the center of the structure ($n\simeq N/2$).
	Good agreement between the upper bound (\ref{eq:IntensityMax_n}) and the numerical results is observed, especially on the left-hand side ($n<N/2$), which indicates that the numerically obtained solutions are indeed the optimal ones.
	The upper bound is evaluated using the optimal reflection coefficient $R=0.219$, which maximises (\ref{eq:IntensityMax}), and which is close to the reflection coefficients (a) $R=0.250$ and (b) $R=0.256$ of the numerically optimized stacks. 
	}
	\label{fig:upperBoundIntensity}
\end{figure}

The above upper bound will be realized only under the condition that the amplitudes $v_n=(A_n,B_n)^T$ inside each layer coincide with the eigenvector $w_n^{(1)}$ or $w_n^{(2)}$ corresponding to the required eigenvalue $\lambda_n^{(1)}$  or $\lambda_n^{(2)}$.
According to (\ref{eq:Eigensys}), both eigenvectors have the property $|A_n|=|B_n|$. Due to flux conservation, see (\ref{eq:fluxConserv}), this property cannot be fulfilled precisely (since $T=0$ implies $A_n=B_n=0$, for all $n$), but asymptotically for large $I_n$. 
Moreover, the requirement that, in this asymptotic limit, the transfer matrix $M_n$ must map the vector $w_n^{(1)}$ onto the vector $w_{n-1}^{(2)}$, or $w_n^{(2)}$ onto $w_{n-1}^{(1)}$, depending on whether $n$ is even or odd, in order to saturate the upper bound, can be used to determine the thickness of the $n$-th layer as $d_n=\lambda_n/4$ (for $N\to\infty$). For finite $N$, the actual, numerically determined optimal thicknesses (see table \ref{tab:thicknesses} below)  may differ from this asymptotic value, especially close to the sample boundaries (i.e. for $n\simeq 1$ and $n\simeq N$, respectively), where the intensities are still small, and close to the sample center, where the transition from exponentially increasing to exponentially decreasing intensities occurs. 

In figure \ref{fig:upperBoundIntensity}, we confirm these considerations by a comparison of the upper bound (\ref{eq:boundright}), (\ref{eq:boundleft}) with numerically optimized solutions of the wave equation (\ref{eq:FullE}). We see that the numerical solutions well reproduce the behaviour predicted by the theoretical upper bound. 
Small deviations are observed on the right-hand side, which originate from the fact that, due to restrictions imposed by the boundary conditions,  the amplitudes $(A_N,B_N)$ in the rightmost layer are not yet close enough to the vector $w_N^{(2)}$ giving rise to the maximum rate of exponential increase.

The exponential behaviour of the upper bound (\ref{eq:IntensityMax}) also explains why the numerical results are rather insensitive to the precise choice of the target function (\ref{eq:target}) of the optimization procedure, see section~\ref{subsec:MethodsOptimization}. 
Indeed, remember that the upper bound was derived by maximizing the background intensity $I_n$, as given by (\ref{eq:Indefinition}),
close to the center of the slab, whereas the numerical optimization maximizes $\gamma$ as defined by equation (\ref{eq:target}). 
Obviously, in both cases, the best strategy is to realize, as closely as possible, the maximum rate of exponential increase given by the upper bound (\ref{eq:IntensityMax}) -- although different choices of the target function may lead to slight changes of the optimal value of $R$, which determines the intensities in the outermost layers.

Therefore, the exponential dependence of the upper bound (\ref{eq:IntensityMax}) for the background intensity suggests a similar behaviour for the optimized value $\gamma^{({\rm max})}(N)$ of the target function $\gamma$: 
\begin{equation}
\gamma^{(\rm max)}(N) \approx \sqrt{\alpha} \; \gamma^{(\rm max)}(N-2) \;\;.
\label{eq:gammaGrow}
\end{equation}
Although (\ref{eq:gammaGrow}) is useful as an approximate analytical benchmark, 
we do not expect it to hold exactly, since $\gamma$ depends on details of the structure --  in particular the thicknesses $d_n$ and the positions of the full intensity profile's minima and maxima with respect to the layers containing the upconverter material -- which are difficult to evaluate analytically.

\subsection{Properties of optimized structures}
\label{subsec:ResultsOptimization}

As already mentioned in the introduction, also Bragg stacks, i.e. photonic structures which consist of layers with periodically alternating thicknesses $d_j=d / ( 4 \sqrt{\eps_j} )$, lead to an enhanced upconversion efficiency \cite{hofmann_upconversion_2016}.
There, the only optimization parameter is the design wavelength $d$ \cite{hofmann_upconversion_2016}. It is therefore to be expected that our present approach with the individual layer thicknesses $d_n$ as tunable control parameters 
enables further improvement, as we demonstrate hereafter.  

\begin{figure} 
	\centering
		\includegraphics[width=0.5\textwidth]{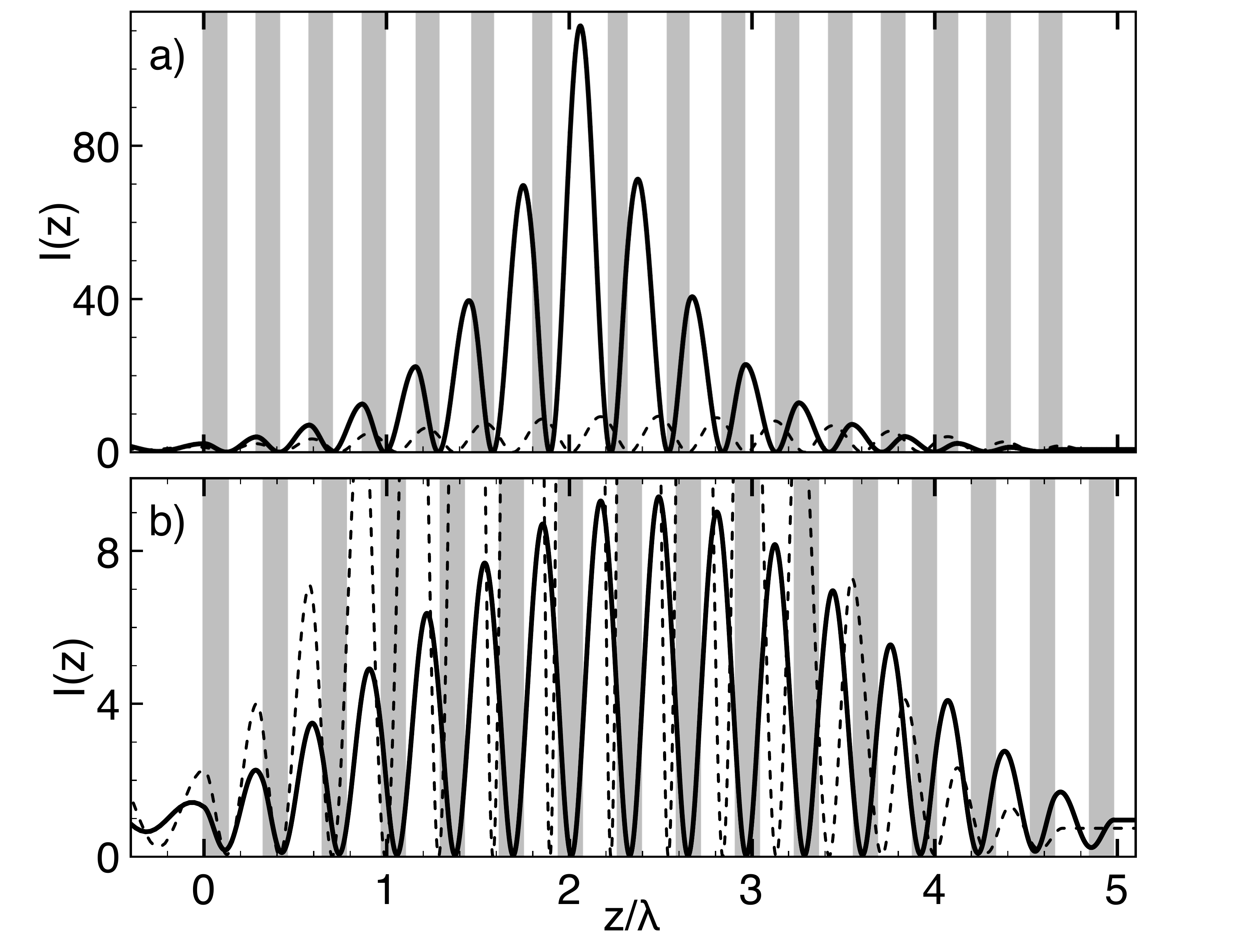}
	\caption{a) Intensity profile $I(z)$ across the photonic structure with individually optimized thicknesses (solid lines), for $N=31$ layers with $\epsilon_{\rm low}=2.25$ and $\epsilon_{\rm high}=4$. The grey vertical bars indicate the layers with higher permittivity
	$\epsilon_{\rm high}=4$. b) Same as a), but for the optimal Bragg stack (with design wavelength $d= 1.107 \lambda$) instead of the structure with individually optimized thicknesses. In the center of the stack, the intensity is less than a $10$th of that achieved in a) (note the different scales of the vertical axes!). To enable a direct comparison between a) and b), the solid lines in a) are drawn as dashed lines in b), and vice versa.
	\label{fig:IntensityPositionN31}}
\end{figure}

\subsubsection{Local intensity profile}
To start with, figure~\ref{fig:IntensityPositionN31} shows the intensity profiles $I(z)$ for a photonic structure with individually optimized thicknesses (a), and for the optimal Bragg stack (b). 
As an example of experimentally realistic parameters \cite{hofmann_upconversion_2016}, we choose $\eps_{0}=\eps_{N+1}=1$, $\eps_{{\rm high}}=2^2$ and $\eps_{{\rm low}}=1.5^2$ for the permittivities of the respective materials, for a structure with $N=31$ layers.
In  figure~\ref{fig:IntensityPositionN31} a), we see that, as a consequence of the individually optimized layer thicknesses - with the upconverting layers' thicknesses increasing from the sample edge to its center (for specific values, see table~\ref{tab:thicknesses} in \ref{sec:dnLists}), the maximum intensity reached within the structure is more than 100 times larger than the incident intensity. 
In the case (b) of the Bragg stack, the intensity is also enhanced, but by less than a factor ten, with an optimized design wave length $d=1.107\lambda$, where $\lambda = 2 \pi c / \omega$ denotes the wavelength of the incident wave in vacuum.
In both, a) and b), $I(z)$ has its maxima within the layers with low permittivity (which contain the upconverter material) and, correspondingly, minima in the layers with higher permittivity.
This is expected as a consequence of optimizing the intensity enhancement factor $\gamma$ inside those layers, which contain the upconverter, see (\ref{eq:target}). This factor results as $\gamma\simeq 30$ and $\gamma\simeq 7$ in a) and b), respectively.

\subsubsection{Enhancement of intensity for increasing number of layers}

\begin{figure}
	\centering
		\includegraphics[width=0.5\textwidth]{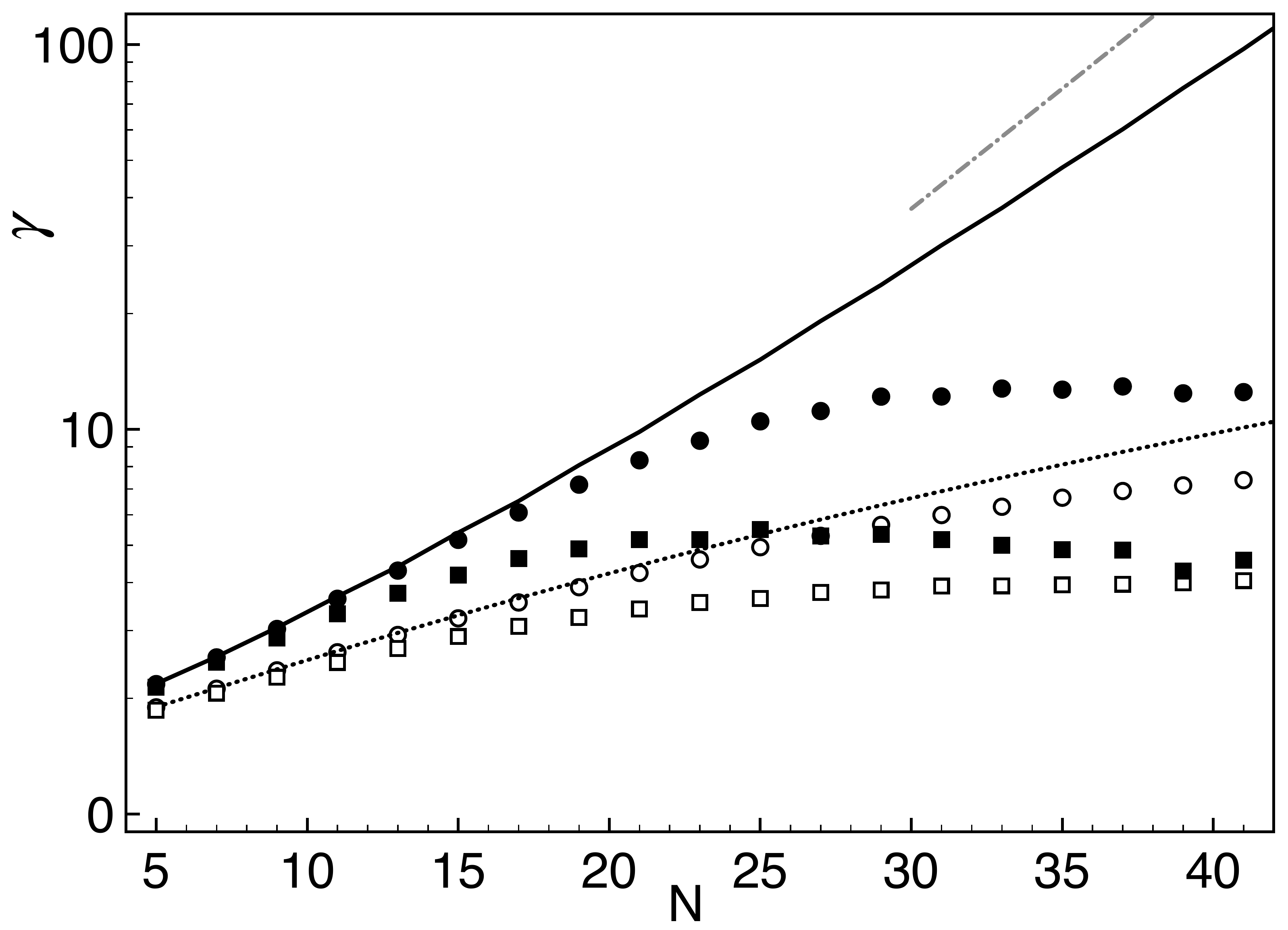}
	\caption{Intensity enhancement factor $\gamma$ for the fully optimized structure (solid line) and for the optimized Bragg stack (dashed line) as a function of the layer number $N$, with $\epsilon_{\rm low}=2.25$ and $\epsilon_{\rm high}=4$. For the fully optimized structure, an exponential increase of $\gamma$ is observed, with a rate which approximately agrees with the analytical benchmark (\ref{eq:gammaGrow}) (grey dash-dotted line).
	Random deviations which account for manufacturing imperfections reduce $\gamma$, especially for the case of the fully optimized structure (filled symbols), where $\gamma$ saturates or even decreases for large $N$, depending on the standard deviation of the errors (circles: $1\%$, squares: $3\%$). However, within a finite range of $N$ the latter retains its advantage over the Bragg stack (open symbols).}
	\label{fig:IntensityvsNlog}
\end{figure}

Let us now examine the scaling behaviour of the intensity enhancement factor $\gamma$ with the number $N$ of layers, see figure~\ref{fig:IntensityvsNlog}. For large $N$, the structure with individually optimized thicknesses (solid line) exhibits an exponential increase, which approximately agrees with the  analytical benchmark (\ref{eq:gammaGrow}) (grey dash-dotted line). But even for small samples with only $N=11$ layers, individual optimization still leads to an improvement by a factor of approximately $1.4$ with respect to the Bragg stack (dashed lines).

Furthermore, figure \ref{fig:IntensityvsNlog} also quantifies the impact of uncontrolled deviations from the optimized structures as unavoidably induced by fabrication errors.
We allow for statistical variations of the $d_n$, which we sample from a Gaussian distribution with standard deviation $\sigma_n$ chosen as a certain fraction ($1\%$ or $3\%$) of the desired mean value $\overline{d_n}$, and determine the intensity enhancement factor $\gamma$ averaged over 10.000 random configurations (symbols).
We see that the structure with individually optimized thicknesses (filled symbols)  is more sensitive against errors than the Bragg stack (open symbols).
This is expected, since, by construction, any deviation from the fully optimized structure decreases $\gamma$, whereas, in case of the Bragg stack,
deviations may also increase $\gamma$, even if reducing it on average.
For the fully optimized structure subject to random errors, the enhancement factor starts to deviate from the exponentially increasing dependence on $N$, as predicted by (\ref{eq:gammaGrow}) at a certain number of layers (which depends on $\sigma_n$), and no longer increases (or even slightly decreases) for larger $N$.
However, even for relatively large errors ($3\%$, filled squares), the fully optimized design retains its advantage over the Bragg stack (open squares) for small layer numbers up to $N\simeq 25$, where the enhancement factor still is, on average, $1.5$ times larger than for a Bragg stack with the same distribution of fabrication errors.

\section{Conclusion}
\label{sec:Conclusion}

We have shown how the individual tuning of the layer thicknesses of a two-component photonic structure allows to push the irradiance concentration in the sample's upconverting sub-volume close to an analytically derived upper bound.

The physical mechanism responsible for this enhancement is the appropriate fine-tuning of interferences between wave amplitudes which are multiply reflected within the multilayer stack. This makes it possible to trap photons of a given wavelength inside the stack, for a time which increases exponentially with the number of layers.
The rate of the exponential increase is proportional to the ratio  $\alpha=\eps_{\rm high}/\eps_{\rm low}$  of the permittivities of the two materials forming the multilayer stack (or, equivalently, to the square of the ratio of their refractive indices).
Random variations around the optimal structure limit this increase, but our optimized structure still performs significantly better than the Bragg stack, for not too large numbers of layers.

We thus conclude that the optimized design of the multilayer structure proposed in the present paper bears a considerable potential for applications, such as upconversion, which benefit from an enhancement of the local field intensity (or, equivalently, the local  irradiance) due to more efficient trapping of the incident photons. While our present paper focuses on the enhancement of the intensity, there are several additional aspects that need to be addressed in future work: indeed, the upconversion luminescence does not only depend on the local irradiance, but, e.g., also on the photonic density of states (which controls the spontaneous emission rates  between the atomic energy levels of the upconverter), and on the efficiency of energy transfer processes relevant for upconversion. The resulting upconversion luminescence can be estimated using a rate equation model  \cite{herter_increased_2013}. Moreover, absorption of the incident photons by the upconverter material, and a finite spectral illumination and absorption bandwidth must be taken into account  in a more complete description, with the aim of achieving a reliable prediction and optimisation of the full upconversion efficiency in experimentally realisable photonic structures under realistic manufacturing constraints.

\section{Acknowledgements}

We thank Clarissa Hofmann and Jan Christoph Goldschmidt for fruitful discussions. 
This research is funded by the Baden-W{\"u}rttemberg Ministry of Science, Research and Arts, the Baden-W{\"u}rttemberg Ministry of Finance and Economy as well as the Headquarters of the Fraunhofer-Gesellschaft in Munich in the project \emph{NaLuWiLeS: Nano-Strukturen zur Lumineszenzverst{\"a}rkung f{\"u}r die Wirkungsgradsteigerung von LEDs und Solarzellen} of the Sustainability Center Freiburg.

\appendix

\section{List of optimized thicknesses} \label{sec:dnLists}

Table~\ref{tab:thicknesses}, lists exemplary structure information on optimized structures, for $N=9,11,19$ and $21$.
Note that most layer thicknesses $d_n$ are close to $1$ (in units of $\lambda_n/4$). Only layers with even $n$, close to the center of the stack
(i.e., $n=4$ for $N=9$ and $N=11$, $n=8$ for $N=19$, and $n=8$ and $10$ for $N=21$) are exceptionally thick. As we have checked, these are the layers where the highest intensities are achieved.

\Table{\label{tab:thicknesses}
Table with 
normalized layer thicknesses $d_n$ for optimized structures of variable layer number $N$. 
Thicknesses are given in units of $\lambda_n/4=\lambda/(4\sqrt{\eps_n})$, $n=1,\ldots ,N $, where $\lambda$ denotes the wavelength of the incident light in vacuum. The permittivities of the layers are $\eps_n=\eps_{\rm high}$ for odd $n$,
and $\eps_n=\eps_{\rm low}$ for even $n$, with $\eps_{\rm high}=4$ and $\eps_{\rm low}=2.25$.}
\br
&\centre{4}{Thickness $d_n$ in units of $\lambda_n/4$}\\
\ns
Layer&\crule{4}\\
\# n & $N=9$ & $N=11$ & $N=19$ & $N=21$ \\
\mr
1  & 1.18957 & 1.12988 & 1.07204 & 1.06752 \\
2  & 1.15274 & 1.03859 & 0.91842 & 0.90985 \\
3  & 1.19511 & 1.09297 & 1.09861 & 1.10123 \\
4  & 1.52053 & 1.62375 & 0.97791 & 0.95032 \\
5  & 1.13942 & 1.11301 & 1.06415 & 1.07680 \\
6  & 0.97623 & 1.12577 & 1.15798 & 1.05228 \\
7  & 1.18056 & 1.16118 & 0.97708 & 1.02027 \\
8  & 0.86353 & 0.90004 & 1.74529 & 1.49702 \\
9  & 1.12043 & 1.18960 & 0.96743 & 0.91514 \\
10  &  & 0.83708 & 1.20913 & 1.60312 \\
11  &  & 1.11609 & 1.06386 & 1.00137 \\
12  &  &  & 0.98526 & 1.08003 \\
13  &  &  & 1.10202 & 1.06578 \\
14  &  &  & 0.91747 & 0.96398 \\
15  &  &  & 1.11970 & 1.09285 \\
16  &  &  & 0.88815 & 0.91942 \\
17  &  &  & 1.12853 & 1.10601 \\
18  &  &  & 0.87421 & 0.89858 \\
19  &  &  & 1.07022 & 1.11274 \\
20  &  &  &  & 0.88835 \\
21  &  &  &  & 1.06078 \\
\br
\end{tabular}
\end{indented}
\end{table}

\section*{References}

\bibliographystyle{iopart-num}

\bibliography{BibPaper1}

\end{document}